\documentclass[twocolumn,nofootinbib,aps,showpacs,jstat,preprintnumbers]{revtex4-1}

\usepackage{graphicx}
\usepackage{dcolumn}
\usepackage{epsfig}
\usepackage{epstopdf}
\usepackage{amssymb}
\usepackage{amsfonts}
\usepackage{xcolor}
\usepackage{float}
\usepackage{amsmath}
\usepackage{hyperref}
\usepackage{physics}
\usepackage{comment}
\usepackage[caption=false]{subfig}
\usepackage{soul}
\usepackage[bottom]{footmisc}

\hypersetup{
	colorlinks = true,
	linkcolor = blue,
	citecolor = blue,
	urlcolor = blue
}
\allowdisplaybreaks
\newcommand{\fig}[1]{Fig.~#1}
\newcommand{\eq}[1]{Eq.~(#1)}
\newcommand{\Sec}[1]{Sec.~#1}
\newcommand{\app}[1]{Appendix #1}
\newcommand{\subfigimg}[3][,]{
	\setbox1=\hbox{\includegraphics[#1]{#3}}
	\leavevmode\rlap{\usebox1}
	\rlap{\hspace*{40pt}\raisebox{\dimexpr\ht1-2.5\baselineskip}{\textbf{#2}}}
	\phantom{\usebox1}
}
\setstcolor{red}

\begin{document}
	
\title{Measurements and analysis of response function of cold atoms in optical molasses}

\author{Subhajit Bhar$^1$}
\email{subhajit@rri.res.in}
\author{Maheswar Swar$^1$}
\author{Urbashi Satpathi$^2$}
\email{urbashi.satpathi@icts.res.in}
\author{Supurna Sinha$^1$}
\author{Rafael D. Sorkin$^{1,3}$}
\author{Saptarishi Chaudhuri$^1$}
\author{Sanjukta Roy$^1$}
\email{sanjukta@rri.res.in}

\affiliation{$^1$ Raman Research Institute, C. V. Raman Avenue, Sadashivanagar, Bangalore-560080, India.}
\affiliation{$^2$International Centre for Theoretical Sciences, Tata Institute of Fundamental Research, Bangalore 560089, India}
\affiliation{$^3$Perimeter Institute for Theoretical Physics, 31 Caroline Street North, Waterloo, ON N2L 2Y5, Canada}

\date{\today}
	
\begin{abstract}		
\noindent We report our experimental measurements and theoretical
analysis of the position response function of a cloud of cold atoms residing in the viscous medium of an optical molasses 
and confined by a magneto-optical trap (MOT).  
We measure the position response function by
applying a transient homogeneous
magnetic field as a perturbing force.  We observe a transition from a damped
oscillatory motion to an over-damped relaxation, stemming from a competition between 
the viscous drag provided by the optical molasses
and the restoring force of the MOT.  
Our observations are in both qualitative and quantitative agreement with
the predictions of 
a 
theoretical model based on the Langevin equation.  
As a consistency check, 
and as a prototype for future experiments, 
we also study 
the free diffusive spreading 
of the atomic cloud 
in our optical molasses 
with the confining magnetic field of the MOT turned off.
We find that the measured value of the diffusion coefficient agrees 
with the value predicted by
our Langevin model, using the damping coefficient. The damping coefficient was deduced from our measurements of the position
response function at the same temperature.
\end{abstract}

\maketitle 
	
\section{Introduction}
	
The response of a physical system to an applied force can
reveal intrinsic characteristics of the system such as electric
polarisability, impedance of an electronic circuit, magnetic
susceptibility and optical conductivity \cite{Kubo1966, Mazenko2006,balescu, Kumar2020, LRT2020}. 
In a similar context, but without the applied force, the
study of diffusive behaviour can provide crucial information regarding
transport properties 
\cite{Barkai2014, Shapiro2010, Davidson2012}.

\par  
In recent years, the diffusion of a Brownian particle in the presence of
quantum zero-point 
fluctuations 
was analysed in
\cite{Sinha1992,Satpathi2017} starting from the fluctuation-dissipation
theorem (FDT) \cite{Kubo1966,balescu}. 
The key input to the analysis presented in \cite{Satpathi2017} is the
position response function that describes how the particle reacts to
an externally applied force.
The specific response function employed in that paper was suggested by
the model of a viscous medium.
In the present work, we study a concrete experimental realization of such a model,
by utilising a three-dimensional configuration of laser beams known as `optical molasses' which
enables cooling as well as viscous confinement of the atomic cloud. 
We find agreement (in a classical regime) with the type of response
function that was assumed in \cite{Satpathi2017}.

Aside from the intrinsic interest of a direct measurement of the
response function, our experiment lays the groundwork for future
experiments that would access the deep quantum regime, where
some of the most interesting effects discussed in \cite{Satpathi2017}
would show up.

\par
In this paper, we demonstrate a method to measure the position response
function of a cold atomic cloud in a MOT by temporarily subjecting it to 
a homogeneous magnetic field 
(transient oscillation method \cite{Kim2005a}).  
We observe
a transition 
from a damped oscillatory motion to an over-damped motion
of the atomic cloud.  This transition stems from a
competition between the reactive spring-like force coming from the
magneto-optical trap and the viscous drag due to the optical
molasses.

\par
By turning off the MOT magnetic field, we are also able to study the
spatial diffusion of the cold atoms in the viscous medium of an optical
molasses, and we verify the Stokes-Einstein-Smoluchowski relation, as
described in more detail in \app \ref{sec:spatial_diffusion}. 

\par
The motion of a Brownian particle can be analysed in terms of either the
Fluctuation-Dissipation theorem  or the Langevin equation. The FDT (which holds both classically and
quantum mechanically) relates the spontaneous position and velocity
fluctuations of a system in thermal equilibrium to its linear response
to an external perturbing force. This allows the spontaneous
fluctuations to be determined from the time-dependent response-function
and vice versa.

\par
The Langevin 
equation\cite{Kubo1966, Mazenko2006, ford, balescu, Hohmann2017, AJP_Volpe, Deng_2007, 2012_Barkai, Graham2000, Majumdar_2015, Vulpiani_2020}
(in its classical, generalised, and quantum forms)
offers a complementary approach which relates the response function
directly to the fluctuating forces that drive the position-fluctuations.

In this paper, we have adopted the Langevin equation as our starting
point, since it enables easy identification of all the forces coming
into play. In applying it to the theoretical analysis of the
dynamics governing the motion of the cold atoms,
we have treated the MOT as an interesting example of an out
of equilibrium system, 
and we have studied it from the point of view of
statistical physics, rather than from the viewpoint of cold atom
experimenters for whom it serves as a valuable and well-documented
source of cold atoms. 
This 
type 
of analysis can be extended to a variety of physical situations where
one is interested in the motion of particles in a viscous medium.

\par 
The paper is organised as follows: In
\Sec{\ref{sec:preparation_detection_cold_atoms}}, we briefly describe
our experimental setup and methods for preparation and detection of the cold atoms. 
\Sec{\ref{sec:Response_Function}} is devoted to the position response function of the cold atomic cloud.  
In \Sec{\ref{sec:Generalized_Langevin_equation}}, we set up the theoretical perspective. 
In \Sec{\ref{sec:motion_cold_atoms}}, we describe our method for measuring the
response function of the cold atoms, and in \Sec{\ref{sec:experimental_results_theory}},
we compare the analytical results with the experimental observations. 
In \Sec{\ref{sec:Conclusion_Outlook}}, we present some concluding
remarks and future perspective.
There are two appendices.  
The first supplements our treatment of the response function in the body of the paper.  
The second presents our results on the spatial diffusion of the cold atoms.

\section{Preparation and detection of cold atoms} \label{sec:preparation_detection_cold_atoms}
	
\begin{widetext}

\begin{figure}
\centering
\includegraphics[scale = 0.4294]{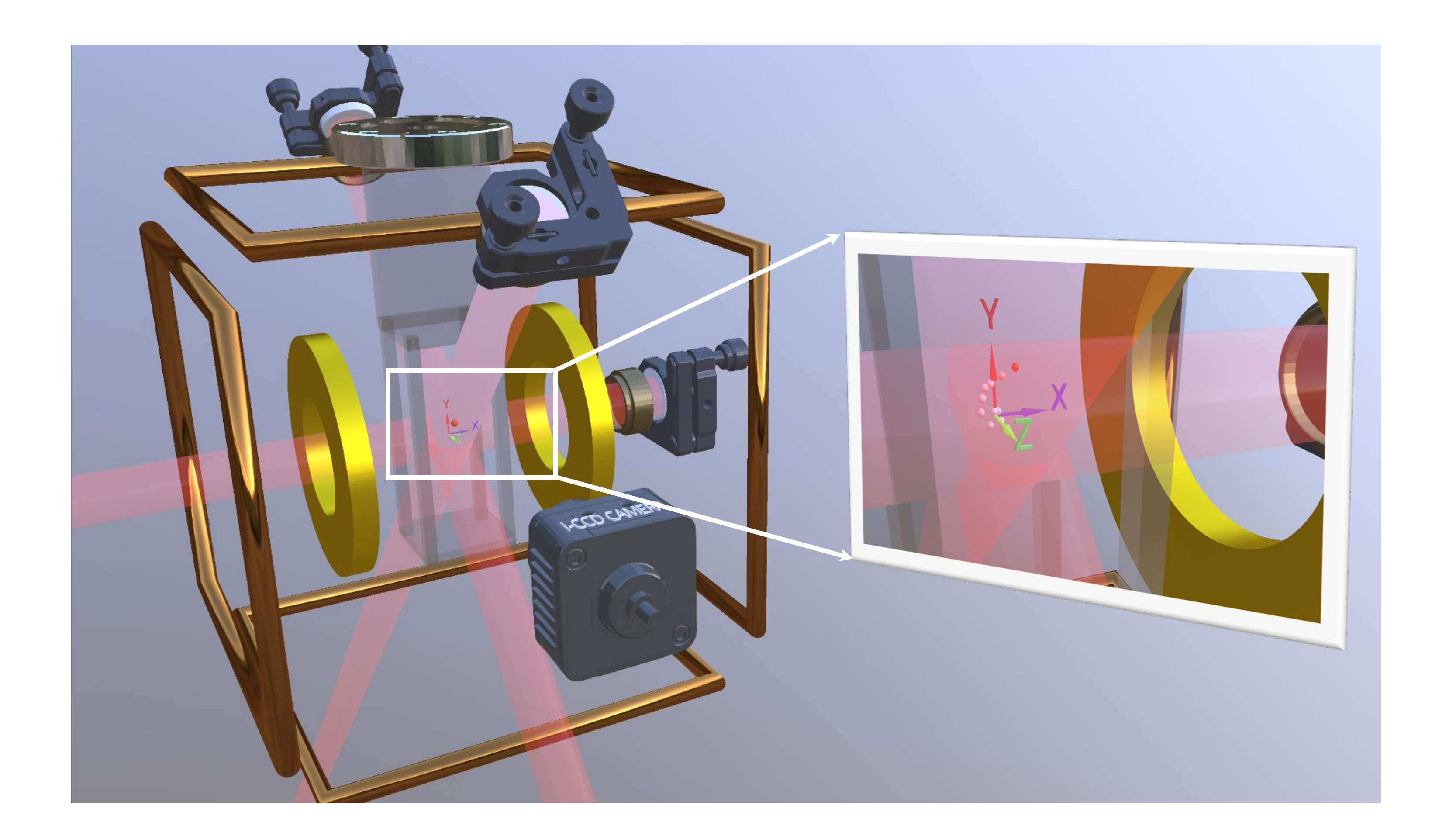}
\caption{Schematic diagram of the experimental setup where a cold atomic
  cloud is produced in a MOT in a glass cell. A magnified view near the
  cold atomic cloud is shown in the inset. The trajectory of the cold
  atomic cloud is shown as a series of atomic clouds at successive
  positions  in the XY plane. The cooling beams are retro-reflected
  using a mirror and a quarter-wave plate. The cylindrical magnetic coils produce the 
  quadrupole magnetic field used for the MOT, and the square
  coils produce the homogeneous magnetic field used in measuring the response function. 
  The detection of the atomic cloud is done by means of absorption imaging using an ICCD camera.} 
\label{fig:experimental_setup}
\end{figure}
\end{widetext}

Our experiment uses a cold atomic cloud of $^{87}$Rb atoms
trapped in a MOT inside an ultra-high vacuum (UHV)  region ($\sim
10^{-11} $ mbar) in a glass cell. 
A schematic diagram of the experimental set-up is shown
in \fig{\ref{fig:experimental_setup}}. 
The MOT 
is 
vapour-loaded from a Rb getter source. 
An external cavity diode laser (ECDL) 
serves
as the cooling laser, 
the laser beam being 12 MHz red-detuned from the 
$5S_{1/2}, F = 2 \rightarrow 5P_{3/2}, F' = 3$ ($D_2$) transition of $^{87}$Rb. 
Another ECDL, the repump laser, is tuned to the transition, 
$5S_{1/2}, F = 1 \rightarrow 5P_{3/2}, F' = 2$, 
and used
to optically pump the atoms back into the cooling cycle. 
This is a standard procedure 
in laser cooling experiments. The detuning and intensity of
the cooling and repump beams are controlled 
by 
acousto-optic modulators (AOM). 
The restoring force required to confine the cold atoms 
is provided by
a pair of current carrying coils 
in a near ideal anti-Helmholtz
configuration.

\par
The fiber coupled  laser beams are expanded to  have a Gaussian waist
diameter of 10 mm and combined in a non-polarizing  cube beam
splitter. Thereafter, the combined cooling and repump beams are split
into three pairs of beams using a combination of half wave plates and
polarizing cube beam splitters. Each of the cooling beams is sent
through the UHV glass cell and retro-reflected via a quarter
waveplate and a mirror. The incoming cooling beams are kept slightly
converging so as to account for the losses in the optical elements and
to ensure that any 
radiation-pressure 
imbalance between the incoming and the
retro-reflected beam is eliminated.

\par  
The cold atoms are detected by a time-of-flight absorption imaging
technique, 
using a short ($\sim$ 100 $\mu$sec) pulse of weak, resonant
linearly polarised 
laser
light
tuned to 
the 
$5S_{1/2}, F = 2 \rightarrow 5P_{3/2}, F' = 3$ transition. 
The shadow cast by the atoms is
imaged onto an ICCD camera with a magnification factor of 0.4. 
In a typical
run of the experiment, we 
trap and cool about 5 $\times$ 10$^7$
atoms at a temperature of around 150 $\mu$K. 

\section{Response Function of the cold atoms}\label{sec:Response_Function}
\subsection{The Langevin Equation}\label{sec:Generalized_Langevin_equation}
	
Our starting point is the Langevin Equation.\footnote
{In this paper we have used the Langevin equation as
the starting point, unlike Ref \cite{Sinha1992,Satpathi2017} where the
FDT was used.  Note, however, that the crucial input to the Langevin equation
is the noise-noise correlation function given in \eq{\ref{eq:noise_properties}}, 
and this rests entirely on the FDT.} 
In its fully quantum mechanical form, it reads 
\begin{equation}
\small
   M\ddot{x} +\int_{-\infty}^{t} dt'\alpha (t-t') \dot{x}(t')+kx = \zeta(t)+f(t) 
\label{QLEq}
\end{equation}
where $M$ is the mass of the particle, $\alpha(t)$ is the
dissipation kernel, 
and  $\zeta(t)$ is the noise 
related to the
dissipation-kernel via the Fluctuation Dissipation Theorem (FDT) \cite{ford}
as follows:
\begin{equation}
\label{eq:noise_properties}
\begin{split}
\small
 \langle \zeta(t)\rangle=&0 \\
 \langle\left\lbrace \zeta(t),\zeta(t')\right\rbrace \rangle\, 
 = \, &\frac{2}{\pi}\int_{0}^{\infty}d\omega\, \hbar\, \omega \, \text{coth}\left( \frac{\hbar\omega}{2k_{B}T}\right)\\
 & \text{Re}[\tilde{\alpha}(\omega)]\text{cos}(\omega(t-t'))
\end{split}
\end{equation}
The position-operator, $x(t)$, of the particle at any time $t$ can be
obtained by solving these equations.

The experiments reported here are at high enough temperatures 
that the noise can be treated classically.
We can therefore take the $\hbar\to0$ limit of the previous equation, to
obtain the noise correlators in their classical form:
\begin{equation}
                \label{eq:noise_classproperties}
                \begin{split}
                        \small
                        \langle \zeta(t)\rangle=&0 \\
                        \langle \zeta(t) \zeta(t') \rangle\, = \, &\frac{2k_BT}{\pi}\int_{0}^{\infty}d\omega\, \text{Re}[\tilde{\alpha}(\omega)]\text{cos}(\omega(t-t'))
                \end{split}
        \end{equation}
(This form of Langevin equation is sometimes termed ``generalized'' to indicate that the dissipation-kernel is not restricted to being a delta-function.)

\par
The last term on the left-hand side of 
\eq{\ref{QLEq}} corresponds to a harmonic force
characterized by a spring constant $k$.  In our present problem $kx$
corresponds to the restoring force of the MOT.  The term $f(t)$
on the right-hand side is a perturbing force, which in this experiment is an additional
magneto-optical force induced by the transient homogeneous magnetic field used to measure the position-response function.

Taking the expectation value of \eq{\ref{QLEq}}, 
and substituting $\langle\zeta(t)\rangle=0$, 
we obtain a deterministic equation for $\langle x(t)\rangle$, 
whose Fourier transform is
\begin{equation}
  \small
  -M \omega^2 \tilde{x}(\omega) - i \omega \tilde{\alpha} (\omega) \tilde{x}(\omega) +k\tilde{x}(\omega)=\tilde{f}(\omega)
  \label{eq: fourier}
\end{equation}
(Here we have used $x$ to denote the mean position of the particle.)
\eq{\ref{eq: fourier}} can be re-expressed as 
	
	\begin{equation}
		\small
		\tilde{x}(\omega) =  \tilde{R}(\omega) \tilde{f}(\omega) ~, 
		\label{eq: resp}
	\end{equation}
	
	\noindent where 
	
	\begin{equation}
		\small
		\tilde{R}(\omega) =\frac{1}{[-M\omega^2 -i\omega \alpha +k]} 
		\label{eq: respexp}
	\end{equation}
is the position response function of the particle
(in this case the cold atomic cloud) in the frequency domain. 

\par
Here we have set $\tilde{\alpha}(\omega)=\alpha$, 
corresponding to the
choice of an Ohmic bath to which the system is coupled. This choice
is motivated by the optical molasses in the cold-atom
experimental setup.  In fact, an Ohmic bath is equivalent
to a force proportional to the velocity with a fixed coefficient of
proportionality or damping coefficient. 
The present experiment serves as a test of this theoretical model of the molasses.    
	
\par 
The position response function in the time domain is given by:
\begin{equation}
  \small
      {R}(t) =\frac{1}{2\pi}\int{\tilde{R}(\omega) e^{-i \omega t} d\omega}
      \label{eq: respexptime}
\end{equation}
The position response function obtained thereby from \eq{\ref{eq: respexp}} for the Ohmic bath is
\begin{equation}
  \small
  R(t)\, = \frac{2}{\alpha_{c}}e^{-\frac{\alpha t}{2M}} \, \text{sinh}\left(\frac{\alpha_{c}t}{2M}\right)
  \label{eq: posrespanalytical}
\end{equation}
where $ \alpha_c =\sqrt{\alpha^2 -4kM} $. 

\par
There are three qualitatively distinct cases. 
For $ \alpha^2 >4kM $, 
$\alpha_c$ is real and 
one gets an overdamped motion of the 
cold atomic cloud.  
For $\alpha^2 =4kM$, 
the motion of the cold atomic cloud is critically damped,
while for $ \alpha^2 <4kM $, 
$\alpha_c$ is imaginary and 
the motion of the cold atomic cloud is a damped oscillation.
For $ k=0 $, \eq{\ref{eq: posrespanalytical}} reduces to the position response function  used in \cite{Satpathi2017}. 
(Note that in the MOT, $k$ is always nonzero due to the presence of the non-zero magnetic field gradient.)

\par
In \Sec\ref{sec:experimental_results_theory}, we 
will
compare the
analytically calculated motion of the 
cold 
atomic cloud that follows from $R(t)$
via equation (\ref{eq:mean_displacement_general}) below
with the experimentally observed oscillatory and damped motions of the cloud.

\par
By taking a time derivative of $R(t)$, one can also get the velocity response function,
\begin{equation}\label{eq: velrespanalytical}
		\small
\dot{R}(t)=\frac{1}{\alpha_{c}M}e^{-\frac{\alpha t}{2M}}\left(\alpha_{c}\text{cosh}\left(\frac{\alpha_{c}t}{2M}\right)-\alpha\,\text{sinh}\left(\frac{\alpha_{c}t}{2M}\right)\right)
\end{equation}

\par
It's an interesting fact that the position response function $R(t)$ can
be inferred directly from the mean velocity induced by a homogeneous force
whose time-dependence is that of a step-function.
By definition, the mean displacement $\langle x(t) \rangle$ is related
to $R(t)$ by the equation, 
\begin{equation}\label{eq:mean_displacement_general}
		\small
		\langle x(t)\rangle= \int_{-\infty}^{t} {R(t-t')f(t')dt'}
\end{equation}
(``linear response theory'' \cite{Sinha1992,Satpathi2017}).
\par
On differentiation, this gives the expectation value $\langle v(t) \rangle$ of the velocity:
\begin{equation}\label{eq:mean_velocity_general}
		\small
		\langle v(t)\rangle= \int_{-\infty}^{t} {\dot{R}(t-t')f(t')dt'}
	\end{equation}
\\
Here $f(t)$ is the external perturbing force, which in our
experiment takes the form of 
a ``top-hat function'', $f(t) = f_0 \,  \theta(t+w) \, \theta(-t)$: 
\begin{equation}
		f(t)=\begin{cases}f_0, &\text{for}\;-w<t<0\\
			0, &\text{for}\; t\leq -w, \;t\geq 0\end{cases}
\end{equation}
(In our experiment $f(t)$ is induced by a bias field.  The temporal profile of
this field, together with an analysis of how $f(t)$ depends on it, is given in
detail in \app \ref{sec:theory_perturbation_force}.)

\par
Substituting into \eq{\ref{eq:mean_velocity_general}},
we get: 
\begin{eqnarray}
		\langle v(t)\rangle &=& f_0\int_{-w}^{0} {\dot{R}(t-t')dt'}\label{eq:mean_velocity_tophat1}\\ &=& -f_0\left(R(t)-R(t+w)\right)\label{eq:mean_velocity_tophat2}
\end{eqnarray}
\begin{equation}\label{eq:resp_vel}
		\small
		R(t)= -\frac{1}{f_0}\langle v(t) \rangle+R(t+w)
	\end{equation}
For $w\rightarrow \infty$, $R(t+w)\rightarrow R(\infty)=0$. 
(This assumes that the MOT is turned on.  When it is turned off and  only the molasses is present, $R(\infty)$ will be nonzero.)
Therefore we get
\begin{equation}\label{eq:resp_final}
		R(t)=- \frac{1}{f_0}\langle v(t) \rangle 
\end{equation}
This simple relationship means that one can measure the position
response function directly, simply by measuring the expectation value of
the velocity.

\subsection{Measurement of the position response function}  \label{sec:motion_cold_atoms}	
The theoretical expression (\ref{eq: posrespanalytical}) for the
position response function of the cold atoms contains two unknown
parameters, the damping-coefficient $\alpha$ and the spring-constant $k$.
In order to test the theoretical model that leads to (\ref{eq: posrespanalytical}), 
and at the same time determine the values of the parameters $\alpha$ and $k$, 
one needs to observe how the cloud of cold atoms moves in response to an external force.

\par
In our experiment we apply a homogeneous magnetic field (bias field), and then
follow the motion of the cloud of cold atoms after the field is switched off. 
We first prepare the laser-cooled
$^{87}\text{Rb}$ atoms in a MOT as described in
\Sec\ref{sec:preparation_detection_cold_atoms}. 
After that, we apply a homogeneous bias field, $B_b$. 
This shifts the trap center to the zero of the new magnetic field. 
The cold atoms experience a force towards the new center, and
equilibrate there within a short interval of time. After 5 sec, we turn
the bias field off, and the cold atoms return to the initial trap
center, following a trajectory from which the position response function
can be inferred. 
To trace the trajectory, we record the position of the cold atoms at
regular intervals of time after turning off the bias field.

\par
\begin{figure}[ht]
  \centering
  \includegraphics[scale = 0.32, trim = {2.75cm 2cm 20cm 0 }, clip]{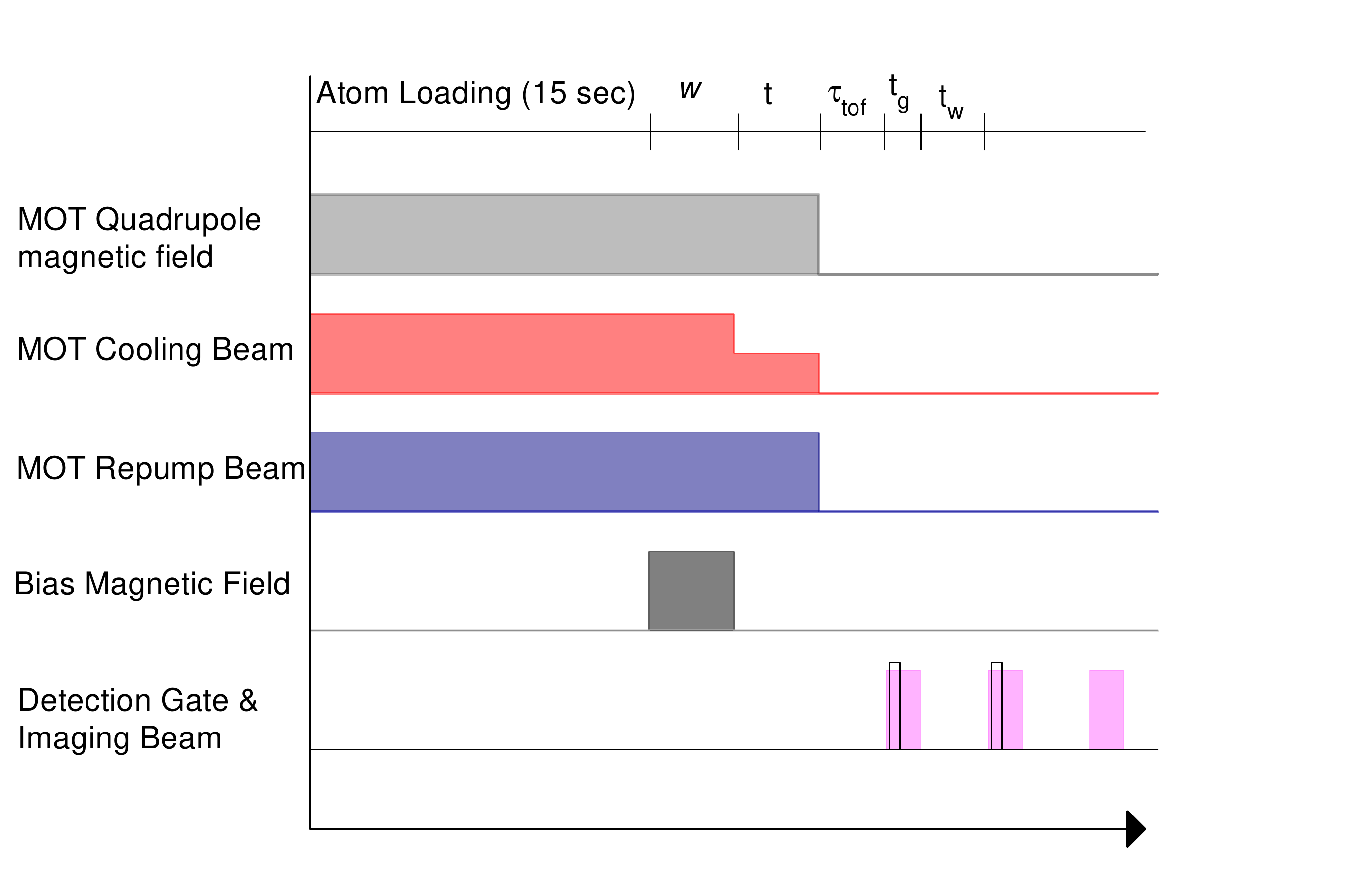}
  \caption{Timing sequence for measuring the response
    function of the cold $^{87}\text{Rb}$ atoms. We 
 prepare 
    the cold atomic cloud by loading the MOT for
    15 sec. Thereafter, we apply a homogeneous bias
    magnetic field for 5 sec ($w$). After the bias field
    is switched off, the cooling beam detuning and
    intensity are changed by a variable amount in order to
    explore a range of values of $\alpha$. The ensuing
    motion of the 
    cloud is monitored via
    time-of-flight absorption imaging. In our experiment,
    time-of-flight($\tau_{tof}$) is 1.2 ms, the detection
    gate time ($t_g$) is 1 ms, and the pulse width of the
    imaging beam is 100 $\mu$s. 
    The time separation
    between successive absorption images ($t_w$) is 1 sec.} 
  \label{fig:responsetimesequence}
\end{figure}

 
\fig{\ref{fig:responsetimesequence}} is a schematic diagram of the 
sequence of events in the experiment. We capture and cool the atoms
in the MOT from a Rb getter source with a loading time of 15 sec. 
The cooling beams, 
having a Gaussian cross-section with a waist size of $10$ mm, 
are red-detuned by $2.2$ $\Gamma$  
from the $5S_{1/2}, F = 2 \rightarrow 5P_{3/2}, F' = 3$ transition,
where $\Gamma = 38.11(6) \times {10}^{6} s^{-1}$ ($2\pi\times  6.065(9)$~MHz)
is the decay rate (natural line-width) of the $^{87}$Rb $D_2$ transition. 
Different values of 
the 
MOT magnetic field gradient 
were used in different sets of measurements. 
For the oscillatory motion shown in \fig{\ref{fig:oscillatory_motion}}, the gradient was $18$ Gauss/cm; 
for the over-damped motion shown in \fig{\ref{fig:overdamped_motion}}, it was $3.5$ Gauss/cm.

\par
After the preparation stage, we apply the bias field for $5$ seconds
(its amplitude being $3$ Gauss for \fig{\ref{fig:oscillatory_motion}}
and $0.6$ Gauss for \fig{\ref{fig:overdamped_motion}}). 
Thereafter, we turn the bias field off and wait for a variable time t, 
after which 
we switch off 
the quadrupole magnetic field and the cooling and repumper laser beams simultaneously, 
and take an absorption image after allowing 
the cloud to move ballistically for a time $\tau_{tof} = 1.2$~ms.  
The mean
position of the cold atomic cloud is inferred 
by
fitting a Gaussian
to the 
column-density 
profile of the 
cloud. 

\subsection{Experimental Results and comparisons with the theory}\label{sec:experimental_results_theory}
\subsubsection{Motion of the cold atoms} \label{sec:position_cold_atoms}
\begin{figure}[ht]
  \centering
  \includegraphics[scale = 0.375]{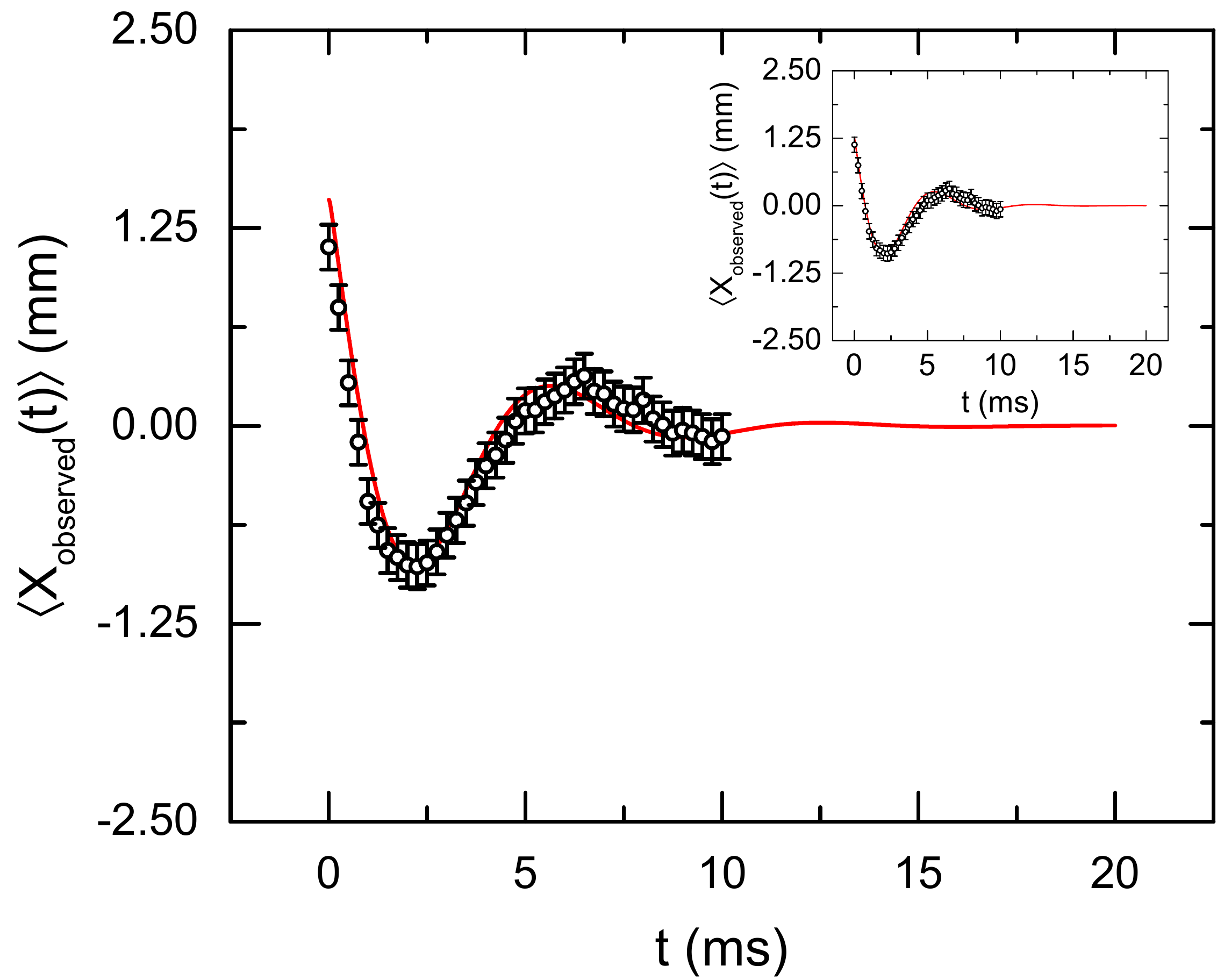}
  \caption{Position of the cold atoms as a function of
    time after the homogeneous bias field is switched off,
    illustrating the underdamped regime. 
    Cooling beam detuning: $-2.2 \,\Gamma$,
    total intensity: $I=16.91\,I_{sat}$; 
    MOT Magnetic field gradient: 18 G/cm;
    bias magnetic field: 3 Gauss with its direction at an
    angle to the image plane. The solid line is the best
    fit between the experimental data and the theoretical
    prediction from \eq{\ref{eq:mean_displacement}} with
    $\alpha \, = \, (1.04 \pm 0.04) \times 10^{-22}$ kg/sec. 
    Inset: A test fit of the data to
    \eq{\ref{eq:damped_harmonic_oscillator}} yielded an
    initial estimate of $\alpha = (1.06 \pm 0.24)\times 10^{-22}$ kg/sec.} 
  \label{fig:oscillatory_motion}
\end{figure}

\begin{figure}[ht]
  \centering
  \includegraphics[scale = 0.375]{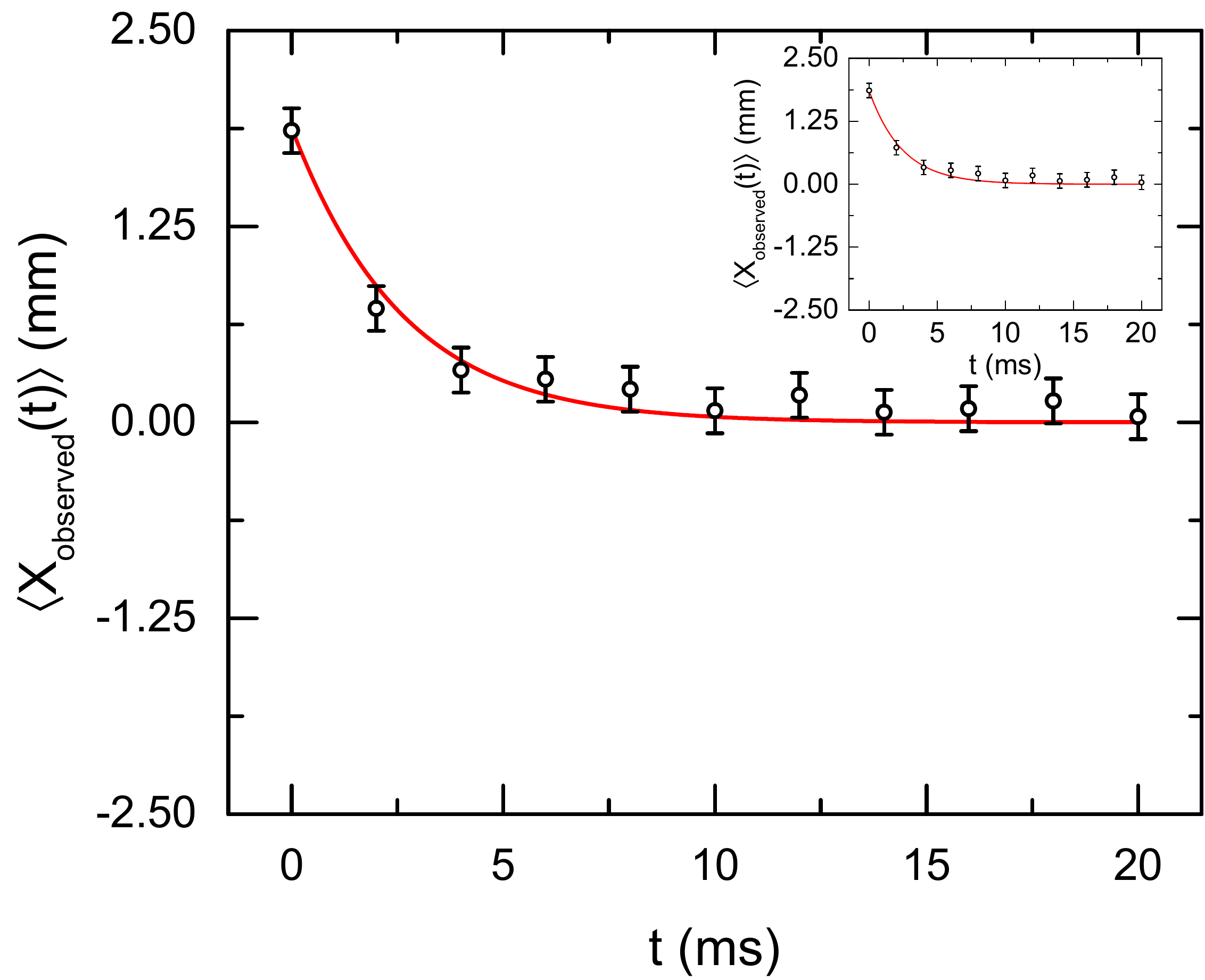}
  \caption{Position of the cold atoms as a function of time after the
    homogeneous bias field is switched off,
    illustrating the overdamped regime. 
    Detuning: $-2.2 \, \Gamma$,  
    total intensity $I$: 9.73 $I_{sat}$; 
    MOT Magnetic field gradient: 3.5 G/cm; 
    bias magnetic field: 0.6 Gauss with its direction along one of
    the Cartesian axes in the image plane. The solid line exhibits the
    best fit between the data and 
    \eq{\ref{eq:mean_displacement}} 
    with $\alpha \, = \, (1.57 \pm 0.46)
    \times 10^{-22}$ kg/sec. 
    Inset: A test fit to
    \eq{\ref{eq:damped_harmonic_oscillator}} yielded an initial
    estimate of $\alpha = (1.58 \pm 0.24) \times 10^{-22}$ kg/sec.} 
  \label{fig:overdamped_motion}
\end{figure}

In our experimental runs, we allow the cloud to move
ballistically for a time  $\tau_{tof} = 1.2$~ms after switching off
the MOT light beams and the quadrupole field
(the bias field having been switched off earlier, of course).
This delay 
lets us acquire the absorption image of the cloud in a magnetic field-free environment. 
However it introduces a small correction to the mean position of the cloud
given by 
\begin{equation}\label{eq:observed_position}
  \langle x_{observed}  \rangle  \, =  \langle x(t) \rangle \,  + \tau_{tof} \, \langle v(t) \rangle 
\end{equation}
The  graphs in \fig{\ref{fig:oscillatory_motion}} and \fig{\ref{fig:overdamped_motion}} show 
the time variation of $\langle x_{observed}\rangle$  
after the bias field is turned off. 
Each data point shown is the average of three experimental runs,   
and the error bar is the standard deviation of the mean position, 
measured as described in \Sec{\ref{sec:motion_cold_atoms}}.

\begin{figure}[ht]
  \centering
  \includegraphics[scale = 0.375 ]{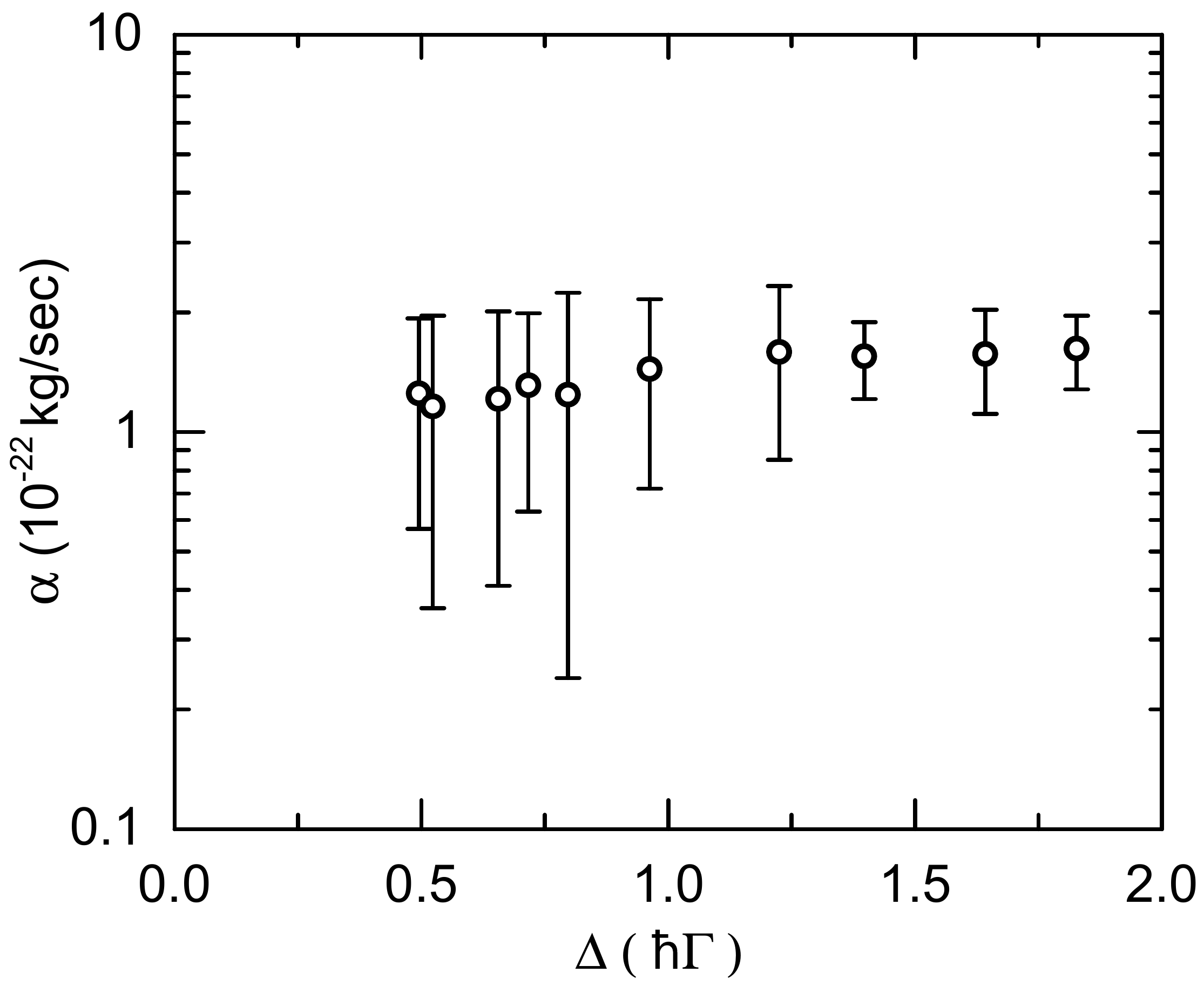}
  \caption{Damping coefficient ($\alpha$) as a function of the light
    shift. MOT magnetic field gradient: 3.5 G/cm. The data was taken in
    the overdamped regime exemplified by
    \fig{\ref{fig:overdamped_motion}}.} 
  \label{fig:alpha_fitted_position_velocity}
\end{figure}

\begin{figure}[ht]
  \centering
  \begin{tabular}{@{}p{\linewidth} @{}}
    \subfigimg[scale = 0.375]{}{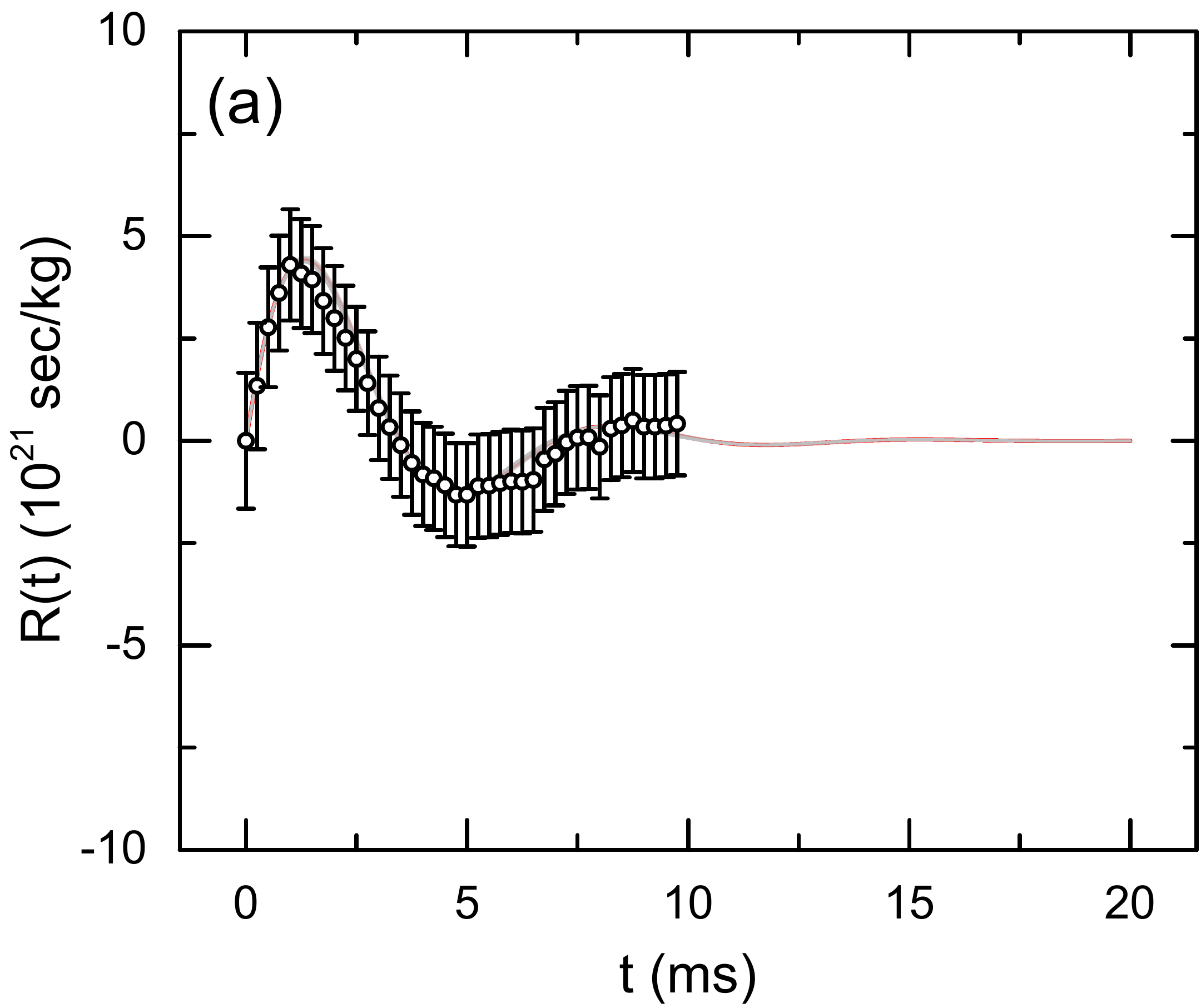} \\[10pt]
    \subfigimg[scale = 0.375]{}{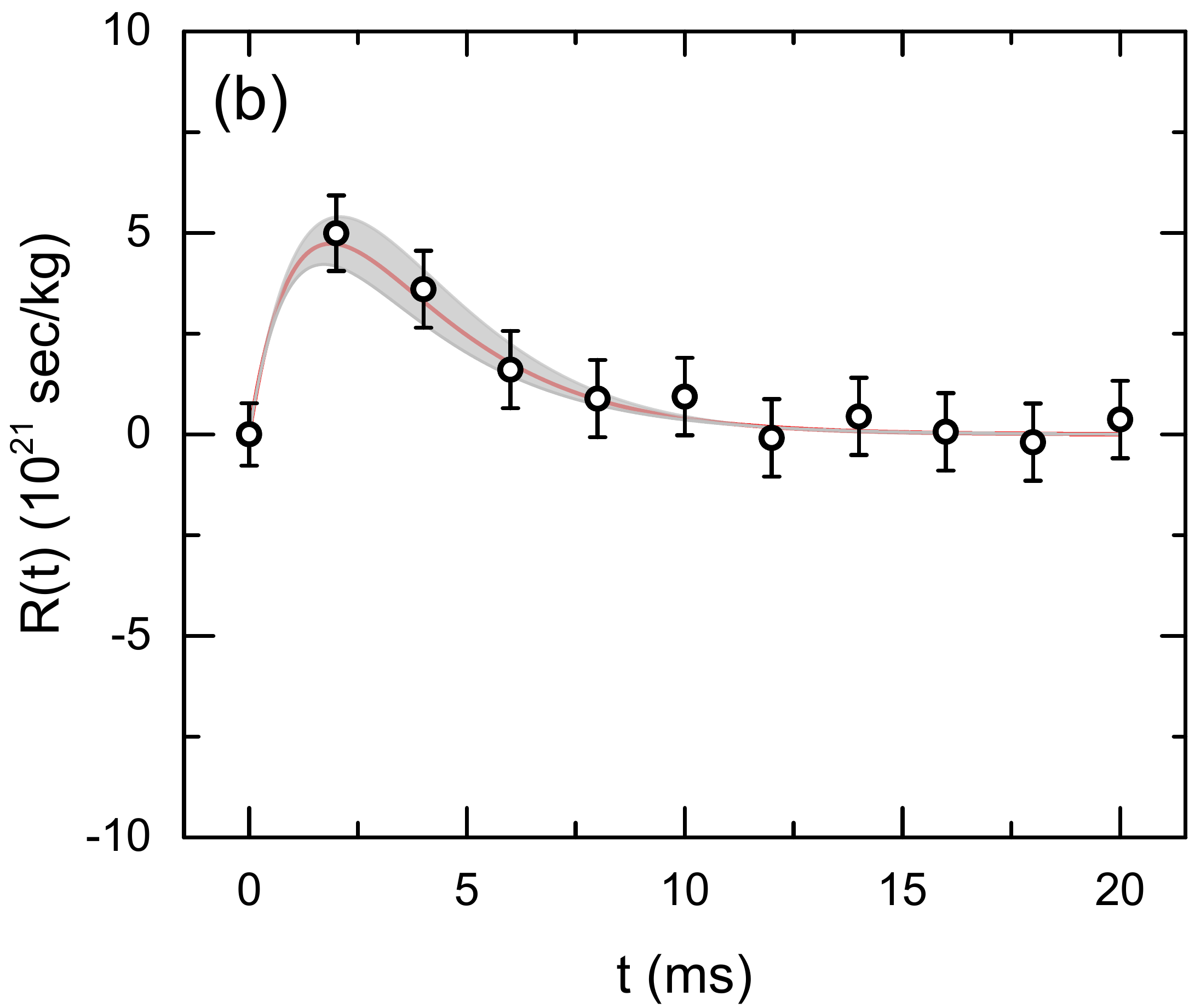}
  \end{tabular}
  \caption{Position response function deduced directly from velocity for (a) oscillatory motion with $\alpha \, = \, ( 1.04 \pm 0.04) \times 10^{-22}  $ kg/sec and (b) 
    damped motion with $\alpha \, = \, (1.57 \pm 0.46) \times 10^{-22} $ kg/sec. In both the graphs, solid lines represent the  
    theoretical prediction of $R(t)$ given in \eq{\ref{eq: posrespanalytical}} 
    and the shaded region shows the $68\%$ confidence band. The
    experimental data points correspond to the scaled velocities
    -$\frac{1}{f_0}\langle v(t) \rangle$ of the cold atoms.} 
  \label{fig:Compare_Response_function}
\end{figure}

\par	
In \fig{\ref{fig:oscillatory_motion}}, we observe an underdamped
oscillatory motion of the cold atomic cloud where the MOT magnetic field
gradient  
is $18$ Gauss/cm and the magnitude of the bias field is $3$ Gauss along
the $x$-direction as shown in \fig{\ref{fig:experimental_setup}}. 
In \fig{\ref{fig:overdamped_motion}}, we observe an over-damped motion of
the cold atomic cloud where the MOT magnetic field gradient is $3.5\,$ Gauss/cm  
and  the magnitude of the bias field is $0.6$ Gauss along the $x$-direction. 

\par	
The theoretical curves shown
in \fig{\ref{fig:oscillatory_motion}} and \fig{\ref{fig:overdamped_motion}}, 
were obtained by fitting the experimental data to 
the prediction 
\eq{\ref{eq:mean_displacement}}
(with due regard to \eq{\ref{eq:observed_position}}).
In these fits, there is only a single fitting parameter:
the damping coefficient $\alpha$.


In the insets to \fig{\ref{fig:oscillatory_motion}} and
\fig{\ref{fig:overdamped_motion}}, we have fitted the experimental data
to the solution of a damped-harmonic oscillator, 
\begin{equation}\label{eq:damped_harmonic_oscillator}
  \langle x(t) \rangle \, = A \, e^{\dfrac{ (-\alpha + \alpha_{c})t}{2M} } \, + \,  B \, e^{\dfrac{ -(\alpha + \alpha_{c})t}{2M} } 
\end{equation}
without assuming anything further about the form of the position response function. 
Here $\alpha_{c}$ is defined as earlier, and 
$M = 1.44316060(11) \times 10^{-25}$ kg
is the mass of the $^{87}Rb$ atom.
The fitting parameters in this case were
$A$, $B$, and $\alpha$.  From these fits, we obtained our initial
estimates of $\alpha$. 

An approach based on a similar 3-parameter fit to the motion of an atom in a MOT was presented 
in \cite{Kim2005a}.  However, while being a correct approximation, it does not capture the 
details of the external perturbing force in their entirety.  In
contrast, our approach based on the 
response function 
can be used for any form of the perturbing force.  Hence it offers a
theoretical model which is versatile and widely applicable for this
class of experiments.

\par
As discussed in \Sec{\ref{sec:Generalized_Langevin_equation}}, 
the cold atomic cloud shows an underdamped oscillatory 
motion or an over-damped motion in response to the applied bias field depending on whether $\alpha^2 < 4kM$ or $\alpha^2 > 4kM$ respectively, 
i.e. whether the restoring force due to the magneto-optical
trapping overwhelms the viscous force due to the optical molasses or vice versa.
As always, $\alpha$ here denotes the 
damping coefficient 
and $k$ the spring constant corresponding to the MOT.
Both $\alpha$ and $k$ can be calculated from 1D Doppler cooling theory \cite{Lett1989, Chang2014} as,	
\begin{align}
  \small 
  \alpha \, & = \, 4 \hbar {\kappa}^2\, s_0 \,
  \dfrac{2\abs{\delta}/\Gamma }{\bigg( 1+ 2 \, s_0 +
    \frac{4\delta^2}{\Gamma^2}\bigg)^2} \label{eq: friction_coefficient}
  \\ 
  k &= g \dfrac{\mu_B \lambda}{h} \, \alpha \,\dfrac{\partial B_m}{\partial x}
\end{align}
where  $\lambda$ is the wavelength and $\kappa = 2\pi/\lambda$ is the wavenumber
of the cooling beams, $\delta$ is the detuning of the cooling beams from the
atomic transition, $\mu_B$ is the Bohr magneton,  
	$\dfrac{\partial B_m}{\partial x}$ is the MOT magnetic field gradient
and $s_0$ is the the saturation parameter of the cooling beams defined as
$I/I_{sat}$ where $I$ is the total intensity of the cooling beams and $I_{sat}$
is the saturation intensity ($I_{sat} = 1.6 \, mW/cm^2$ for $^{87}$Rb $5S_{1/2}, F = 2 \rightarrow 5P_{3/2}, F' = 3$ transition and $\sigma^{\pm}$ polarised
light).  
Hence, the 
damping coefficient 
$\alpha$ depends on the detuning and intensity of
the cooling beams of the MOT,
while
the spring constant $k$  has an additional dependence on the magnetic field gradient. 

\par	
Using the simplest possible assumption that the
        fluorescence from the trapped atoms accurately gives the damping
        co-efficient in our fitting algorithm described above, we obtain a
        normalised mean square residual of 8.2\% and 5\% for the data presented
        in \fig{\ref{fig:oscillatory_motion}} and
        \fig{\ref{fig:overdamped_motion}} respectively. However, in the presence
        of a gradient magnetic field in the MOT and for a Gaussian atom number
        spatial distribution in the atomic cloud and a Gaussian spatial
        intensity profile of the cooling laser beams, this simple assumption is
        likely to be inaccurate. Therefore, we kept $\alpha$ to be a free
        fitting parameter and obtained a normalised mean square residual to be
        2.1\% and 2.6\% for the data presented in
        \fig{\ref{fig:oscillatory_motion}} and \fig{\ref{fig:overdamped_motion}}
        respectively. This indicates that while the fluorescence measurements
        can give a reasonable estimate of the damping coefficient of cold atoms
        in the MOT, more accurate values of the damping coefficient can be found
        using experimental measurements which is modelled well using our
        theoretical description presented in this paper.

\subsubsection{Estimation of the damping coefficient ($\alpha$) in the MOT}
	
In \fig{\ref{fig:alpha_fitted_position_velocity}}, the damping coefficients
($\alpha$) obtained from fitting the experimental data with the analytical
expression given in \eq{\ref{eq:mean_displacement}} and
\eq{\ref{eq:velocity_analytical}} are plotted against the light shifts, where
the light shift ($\Delta$) is given by:    
	
\begin{equation}\label{eq:light_shift}
		\small
		\Delta  \, =  \hbar \, \abs{\delta} \,   \, \dfrac{I/I_{sat}}{1 \, + \, 4 \delta^2 / \Gamma^2}
\end{equation}
	
\noindent where $\delta$ is the detuning of the cooling beam from the atomic transition and $\Gamma$ is the natural linewidth of the atomic transition having transition wavelength $\simeq$ 780 nm.

\subsubsection{Position response function from velocity}
In \fig{\ref{fig:Compare_Response_function}a} and \fig{\ref{fig:Compare_Response_function}b}, 
we show comparisons between the theoretically obtained position response
functions given in \eq{\ref{eq: posrespanalytical}} (solid lines) and
the experimentally  
obtained scaled velocities  -$\frac{1}{f_0}\langle v(t) \rangle$  (circle with error bars) for the motion  of the atomic clouds given in 
\fig{\ref{fig:oscillatory_motion}} (oscillatory motion) and \fig{\ref{fig:overdamped_motion}} (damped motion).
Note that the scaled velocity data agrees very well with the curves for the response functions, confirming 
\eq{\ref{eq:resp_final}}, which is indeed a very good approximation to the exact response function (in other words, the top hat 
	function approximates the exact bias field and in turn the perturbing force well).

\smallskip
\par
As we vary the molasses parameters and the MOT's magnetic field-gradient in the
experiment, we observe both oscillatory and monotonic motions of the cloud's
centroid $\langle x(t)\rangle$, indicating a transition from an underdamped to
an over-damped regime.
We did not attempt to explore all the parameters (intensity, detuning, magnetic field
gradient) in sufficient detail to pin down the exact transition point between the two
regimes.  Nevertheless, in the reasonably large parameter space that we have explored, 
the two regimes appear clearly, as does more generally the systematic variation of the response function with the
experimental parameters of the MOT.

\section{Conclusion and Outlook} \label{sec:Conclusion_Outlook}
In this work we have measured the position
response function of the cold atoms in a MOT 
by subjecting them to a transient
homogeneous magnetic field. We have tested theoretical predictions
regarding the nature of the response function, 
and we
have done extensive theoretical analysis and numerical modelling of our
experimental observations.  

One of the significant outcomes of the study has been the verification of the
functional form of the position response function which was used as input to a
recent theoretical study\cite{Satpathi2017} of diffusion, not only in the
classical domain dominated by thermal fluctuations, but also in the
still-to-be-explored quantum domain where zero point fluctuations
are the main driver of the diffusion \cite{footnote}.

\par
Our study has led to an interesting experimental observation of a transition from an oscillatory to an 
over-damped behaviour of the response function as a result of a competition between elastic and dissipative 
effects. We find a good agreement between our experimental measurements and the theoretical
model of a particle moving in a viscous medium and confined by a
harmonic-oscillator potential. These measurements can be readily extended to 
lighter atomic species compared to Rb such as Na and K so as to access a larger range of parameter 
space to observe a smooth transition of the response function from an under-damped to an over-damped behaviour. 
 
\par
We also studied the spatial diffusion of the cold atoms in the optical
molasses (\app \ref{sec:spatial_diffusion}), observing a behaviour which
is consistent with a theoretical model based on the Langevin
equation. 
In particular, the measured value of the diffusion coefficient
agreed with the value predicted by the Langevin model, using the damping
coefficient deduced from our measurements of the position response
function at the same temperature. 

\par
One novelty of our theoretical analysis is the observation that the position
response function can be obtained directly from the velocity
(\eq{\ref{eq:resp_final}}) if the temporal variation of the perturbing force is
a step function.  This is confirmed by our experiment.

\par
Our theoretical analysis also points out that in the MOT where the magnetic field is
linearly proportional to the distance from the centre, the magneto-optical force
can be written as the gradient of the square of the local magnitude of the
magnetic field as shown in \eq{\ref{eq:MOT_Force}} of the first Appendix.
This relationship simplified the theoretical modelling of the perturbing force in our experiment as seen
in \eq{\ref{eq:MOT_bias_Force}}.  
	
\par
Our study provides a general framework to analyse the motion of a particle in
optical molasses combined with a restoring force, such as in a MOT, ion-traps in
the presence of cooling laser beams \cite{Radium2019}, or ultra-cold atoms in optical lattices in
the presence of additional optical molasses \cite{Bloch2010, Greiner2009}.  These and other similar
experimental systems are of current interest in the context of quantum
technology devices \cite{Amico_2017}.  
This study paves the way for exploring spatial diffusion of
ultra-cold atoms in the quantum regime 
where zero point fluctuations dominate over thermal ones 
\cite{Sinha1992,Satpathi2017,Das2020}.

\par
The central questions addressed in this paper are rooted in
non-equilibrium statistical mechanics, and the fact that we address them using the tools
of cold atom physics makes this study inherently interdisciplinary in nature.
In future we intend to experimentally measure and analyse the zero point fluctuation driven diffusion in the quantum domain that has been predicted in 
\cite{Sinha1992,Satpathi2017}.  In that context we will expand our perspective beyond the classical Langevin Equation to a fully 
quantum mechanical formulation (quantum Langevin equation).

\begin{acknowledgements}
This work was partially supported by the Ministry of Electronics and Information Technology (MeitY), Government of India, through the Center for Excellence in Quantum Technology, under Grant4(7)/2020-ITEA. S.R acknowledges funding from the Department of Science and Technology, India, via the WOS-A project grant no. SR/WOS-A/PM-59/2019. This research was supported in part by Perimeter Institute for Theoretical Physics.  Research at Perimeter Institute is supported in part by the Government of Canada through the Department of Innovation, Science and Economic Development Canada and by the Province of Ontario through the Ministry of Colleges and Universities. We acknowledge Hema Ramachandran, Meena M. S., Priyanka G. L. and RRI mechanical workshop for the instruments and assistance with the experiments. 
	\end{acknowledgements}
	
\begin{appendix} 
\renewcommand{\thesection}{\Alph{section}}
\section{Theoretical Modelling: Response function of the cold atoms}\label{sec:theoretical_modelling}
\subsection{Perturbing force on cold atoms subjected to a transient homogeneous magnetic field}\label{sec:theory_perturbation_force}
\begin{figure}[ht]
\centering
\includegraphics[scale= 0.36, trim = 1.75cm 1cm 0 0,clip ]{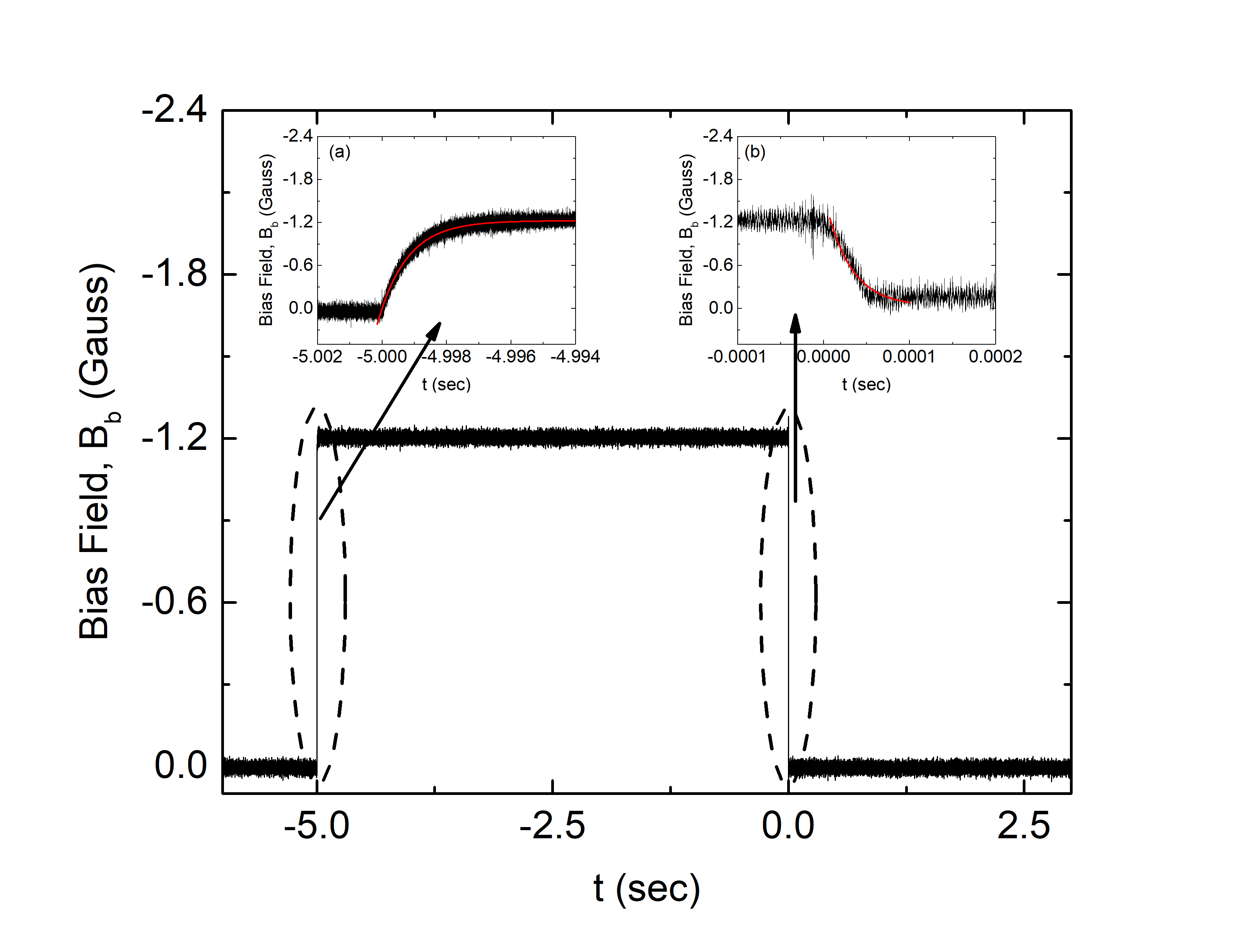}
\caption{Temporal profile of the bias field.  The black solid line is the
  experimental data recorded using a pick-up coil. Insets (a) and (b) show the
  growth and the decay, respectively, of the bias field as a function of
  time. After fitting the data using \eq{\ref{eq:bias_field}} we obtain $ \tau_1
  = (912 \pm 0.37) \, \mu sec \, \text{ and } \, \tau_2 = (29.6 \pm 0.056 ) \,
  \mu sec $.}
\label{fig:biasfield}
\end{figure}

The temporal profile of the bias field used in our experiments  is shown in
\fig{\ref{fig:biasfield}}. We fit this profile with the following equation: 
\begin{equation}
\small
B_{b}(t)
\begin{cases}
=0 & \text{if $t \le - w$} \\[15pt]
=B_0  \Bigg(1-e^{-\frac{t+\text{$w$}}{\tau_1}} \Bigg)  & \text{if $-w\leq t \leq 0$} \\[15pt]
=B_0  \Bigg(1-e^{-\frac{\text{$w$}}{\tau_1}} \Bigg) \, e^{-\frac{t}{\tau_2}} & \text{if $t\geq 0 $}\\
\simeq B_0 \, e^{-\frac{t}{\tau_2}} &( \text{for $w >> \tau_1$})
\end{cases}
\label{eq:bias_field}
\end{equation}
\noindent where
$B_0$ is the magnitude and $w$ is the pulse width of the bias field,
and where $\tau_1 \, \text{ and } \, \tau_2$ are the rise time and fall time of the bias field. 
In our experiment $ \tau_1 \, \text{ and } \, \tau_2 $ 
are  $912 \,\, \mu sec  \text{ and } \,\, 29.6  \, \mu sec$ 
respectively.  The approximation done in the last line of \eq{\ref{eq:bias_field}} is due
to the fact that the time duration of the bias field ($w $ = 5 sec) is
much larger than the $912 \,\,\mu sec$ rise time of the bias field. 
The exact values of $\tau_1$ and $\tau_2$ depend on the design
details of the fast switching circuit for the magnetic field coils in
Helmholtz configuration producing the bias field \cite{fast_switching}. 
It is important to have a fast `switching off' of
the magnetic field so as to ensure that the measurements taken after
switching off the magnetic field are not significantly affected by the
time-constant $\tau_2$.  
In any case, we incorporate the effect of $ \tau_1  $ and $ \tau_2 $ on the motion of the atoms in our theoretical model. 

\par 
In a MOT, the $x$-component of the force 
on the cold atoms, 
which in \cite{FootBook} is expressed
in terms of $x\,\partial{B_m}/\partial{x}$, 
can be recast as follows to show that 
the squared $B$-field acts like a potential energy for the atoms:

\begin{equation}\label{eq:MOT_Force}
			\begin{split}
				\small
				F_{MOT} \, &=  - \alpha v - \, g \, \dfrac{\mu_B \lambda	}{h} \, \alpha \, x \,   \dfrac{\partial B_m}{\partial x} \\[5pt]
				&=  - \alpha v - \, g \, \dfrac{\mu_B \lambda	}{h} \, \alpha \, \dfrac{1}{2C_m} \,  \dfrac{\partial B_m^2}{\partial x}
			\end{split}
\end{equation}

Here,
 $g=g_{F'} m_{F'} - g_{F} m_{F} $ 
 for transition between the hyperfine levels $|F,m_F\rangle$ and $|F',m_F'\rangle$, 
 $\mu_B$ is the Bohr magneton, 
$\lambda$ is the wavelength of the cooling beams, 
$h$ is the Planck's constant,
and $\alpha$ is the damping coefficient.
In the second line we have used that
$B_m = C_m x$ with  $C_m$ a constant,
which implies that 
$$x\dfrac{\partial B_m}{\partial x}\,=\,\dfrac{B_m}{C_m} \dfrac{\partial B_m}{\partial x} = \dfrac{1}{2C_m} \dfrac{\partial B_m^2}{\partial x}$$.
\medskip
\par
In the presence of an
additional bias field ($ B_b $) along the negative $x$-direction, the force on an
atom is
given by (\ref{eq:MOT_Force}) with the bias field added to $B_m$:
\begin{equation}\label{eq:MOT_bias_Force}
  \begin{split}
   \small
  F_{net}  \, &=  - \alpha v - \, g \, \dfrac{\mu_B \lambda	}{h} \, \alpha \, \dfrac{1}{2C_m} \,  \dfrac{\partial (B_m - B_b)^2}{\partial x} \\[5pt]
  &=  - \alpha v - \, g \, \dfrac{\mu_B \lambda	}{h} \, \alpha \, \dfrac{1}{2C_m} \,  \bigg( \dfrac{\partial B_m^2 }{\partial x} - \,  2 C_m B_b  \bigg) \\[5pt]
  &= F_{MOT} \, +  \, f(t) 
  \end{split}
\end{equation}
where we used that $\dfrac{\partial B_b }{\partial x} \, = \, 0$. 
Therefore
\begin{equation}\label{eq:perturbation_force}
			f(t)=   g \, \dfrac{\mu_B \lambda}{h} \, \alpha \, B_{b}
		\end{equation}
		
\subsection{Mean displacement of the cold atoms} \label{sec:theory_mean_displacement}
Using the expression for the position response function in
                \eq{\ref{eq: posrespanalytical}} and the perturbing force in
                \eq{\ref{eq:perturbation_force}}, we get
\begin{equation}\label{eq:mean_displacement}
			\small
			\langle x(t)\rangle=A e^{\dfrac{\left(-\alpha+\alpha_{c}\right)t}{2M}}+B e^{\dfrac{-\left(\alpha+\alpha_{c}\right)t}{2M}}+Ce^{\dfrac{-t}{\tau_{2}}}
		\end{equation}
where,
\begin{eqnarray}
A&=&-\frac{2Mf_0}{\alpha_c}\left[\frac{\left(e^{\frac{\left(-\alpha+\alpha_{c}\right)w}{2M}}-1\right)}
  {\alpha-\alpha_{c}}-\frac{\tau_1\left(e^{\frac{\left(-\alpha+\alpha_{c}\right)w}{2M}}-e^{\frac{-w}{\tau_1}}\right)}{\alpha\tau_1 -2M-\alpha_c \tau_1}\right.\nonumber\\
  &&\left.+\dfrac{\tau_2\left(1-e^{\frac{-w}{\tau_1}}\right)}{\alpha \tau_2 -2M -\alpha_c \tau_2}\right]\label{eq:A}\\
  B&=&\frac{2Mf_0}{\alpha_c}\left[\frac{\left(e^{\frac{-\left(\alpha+\alpha_{c}\right)w}{2M}}-1\right)}{\alpha+\alpha_{c}}-\frac{\tau_1\left(e^{\frac{-\left(\alpha+\alpha_{c}\right)w}{2M}}-e^{\frac{-w}{\tau_1}}\right)}{\alpha\tau_1 -2M+\alpha_c \tau_1}\right.\nonumber\\&&\left.+\frac{\tau_2\left(1-e^{\frac{-w}{\tau_1}}\right)}{\alpha \tau_2 -2M +\alpha_c \tau_2}\right]
  \label{eq:B}\\
  C&=&\frac{4M \tau_{2}^{2} f_0\left(1-e^{\frac{-w}{\tau_1}}\right)}{\left(4M^2+\tau_{2}^{2}(\alpha^2 -\alpha_{c}^2)-4M\alpha\tau_{2}\right)}\label{eq:C}
\end{eqnarray}
Here, $f_0=g \, \dfrac{\mu_B \lambda	}{h} \, \alpha \, B_{0}$ from \eq{\ref{eq:perturbation_force}}.
	
The mean velocity can also be obtained by taking a time derivative of $ \langle x(t)\rangle $ in \eq{\ref{eq:mean_displacement}},
\begin{eqnarray}
\langle v(t)\rangle&=&\frac{(\alpha_c -\alpha)A}{2M} e^{\dfrac{\left(-\alpha+\alpha_{c}\right)t}{2M}}-\frac{(\alpha +\alpha_c)B}{2M}\nonumber\\
    && e^{\dfrac{-\left(\alpha+\alpha_{c}\right)t}{2M}}-\frac{C}{\tau_2}e^{\dfrac{-t}{\tau_{2}}}
			\label{eq:velocity_analytical}
		\end{eqnarray}
Note that for negligible $\tau_1, \tau_2$ and for infinite width, i.e. $w\rightarrow \infty$, using \eq{\ref{eq:A}}, \eq{\ref{eq:B}} and \eq{\ref{eq:C}},
 $$\frac{(\alpha_c -\alpha)A}{2M}\rightarrow -\frac{f_0}{\alpha_c}, \frac{(\alpha +\alpha_c)B}{2M}\rightarrow \, - \frac{f_0}{\alpha_c}, \frac{C}{\tau_2}\rightarrow 0$$
then, using \eq{\ref{eq:velocity_analytical}} 
$\langle v(t)\rangle \rightarrow -f_0 R(t)$  ($R(t)$ is given by
                \eq{\ref{eq: posrespanalytical}})  and thus 
                \eq{\ref{eq:resp_final}} is satisfied.

\section{Spatial diffusion of cold atoms}\label{sec:Diffusive_Motion}\label{sec:spatial_diffusion}
We study the diffusive behaviour of the cold atoms in the viscous medium
provided by 
our 
optical molasses, exploring 
different temperatures as the atomic cloud is cooled to lower temperatures via sub-Doppler cooling.

\par
When the restoring force produced by the MOT magnetic field is absent,
we are in the $k=0$ regime of the Langevin equation which defines our
theoretical model. We continue to assume that the dissipation kernel
$\alpha(t)$ is simply a delta-function in time, or equivalently that the
force exerted on an atom by the optical molasses is  
 \begin{equation} \label{eq: force_equation}
		F_{OM} \, = \,  - \alpha v 
\end{equation}
where v is the velocity of the atom. For consistency, one would hope
that essentially the same value of $\alpha$ would explain both the
position response function studied 
above 
and the diffusive spreading
studied here. 

\par
As is well known, the mean-square distance traveled by the diffusing
atom can be determined from the Langevin equation together with the
noise-correlator, i.e. from 
equations (\ref{QLEq}) and (\ref{eq:noise_classproperties}).
The predicted time-dependence of the spreading
depends on how the observation time $\tau$ compares with the
``relaxation time'' $M/\alpha$. When $\tau \gg M/\alpha$ one finds the
familiar Brownian motion, with 
diffusion coefficient $D$ given by
the 
Stokes-Einstein-Smoluchowski 
relation:
\begin{equation} \label{eq: diffusion coefficient}
  D = \frac{k_B T}{\alpha}
\end{equation}
where $k_B$ is the Boltzmann constant and $T$ is the
temperature of the cloud. 
However, for $\tau \ll M/\alpha$
one finds that the mean-square distance travelled grows like $t^3$
rather than $t$. 
In our experiment, the observation time of 20 ms is about
20 times bigger than $M/\alpha\sim 1$~ms. Although this is not
enormously greater than unity, it seems sufficiently big for us to
ignore the short-time crossover to $t^3$ spreading. We have therefore
fitted 
the data under the assumption that we are in the regime of
Brownian motion (Wiener process). 
\begin{figure}[ht]
  \centering
  \includegraphics[scale=0.32]{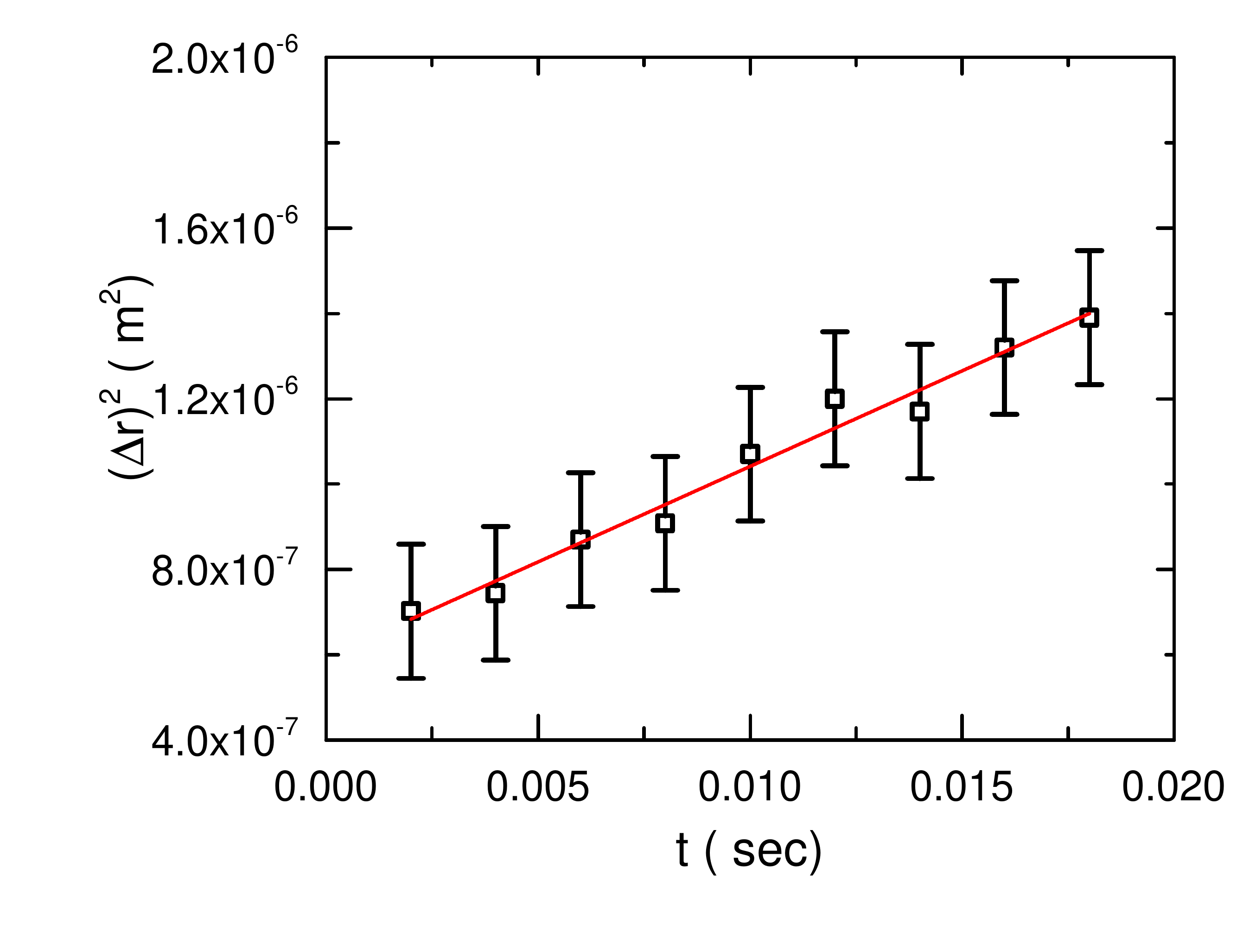}
  \caption{The plot shows the atomic cloud size expanding in an optical molasses 
    at a temperature of around $120\mu K$~. The solid line is a fit to the
    experimental data using  \eq{\ref{eq: experimental diffusion coefficient}}  .}  
  \label{fig:diffusion_coefficient}
\end{figure}

\par 
To observe the diffusive spreading of the atoms in the cold atomic cloud, we first loaded the MOT from the background Rb vapour.
Thereafter, the MOT magnetic field was switched off, and the 
cloud was allowed to diffuse in the presence of the cooling laser beams
forming the optical molasses, but still in the absence of the MOT
magnetic field. 

\par
In the Brownian motion approximation, an atomic cloud of initial size of
$\eval{(\Delta r)^2}_{t = 0}$ expands to a size of $\eval{(\Delta
  r)^2}_{t = \tau}$ in time $\tau$ according to the relation: 
	\begin{equation}\label{eq: experimental diffusion coefficient}
	\small
	\eval{(\Delta r)^2}_{t = \tau}  \,=   \eval{(\Delta r)^2}_{t = 0} \,  + \, 4 \, D\, \tau,
\end{equation}
where, $\Delta r$ is the rms width of the cold atomic cloud.

\par
We obtained 
${(\Delta r)}^2$ directly from the column density profile of the
absorption image at time $t$. 
In other similar experiments \cite{Hodapp1995}, the density profile was
fitted to a Gaussian distribution, whereas the $(\Delta r)^2$ shown in
\fig{\ref{fig:diffusion_coefficient}} was obtained directly from the
absorption images without assuming Gaussianity. This additional
generality could become important in the quantum regime of logarithmic
spreading, for which the analysis of \cite{Satpathi2017} furnishes $
(\Delta r)^2$ but not the full probability distribution of $ \Delta
r$. (We know of no proof that the latter will be Gaussian when the
diffusion is not classical.) 

\par
\eq{\ref{eq: diffusion coefficient}} 
relates the damping coefficient
$\alpha$ to the diffusion coefficient $D$ and thereby allows us to check
for consistency between our direct measurement of D
(\fig{\ref{fig:diffusion_coefficient}}) and the value of the $\alpha$
deduced from our earlier measurements of the position response
function. For a temperature of around $120 \mu K$ of the cold atomic
cloud, the diffusion coefficient obtained from the measurement of the
diffusive spreading of the atomic cloud 
was  
$(1.01 \pm 0.15) \times 10^{-5}$ m$^{2}$/s 
yielding a value of $(1.58 \pm 0.25) \times 10^{-22}$~kg/s for
$\alpha$. 
For the same temperature, the value of $\alpha$
obtained from the measurement of the position response function was
$(1.57 \pm 0.46)\times 10^{-22}$ kg/s.
The agreement could not be better.

                
\end{appendix}

\newpage	
	
\bibliographystyle{apsrev4-1} 
\bibliography{ref}
	
\end{document}